\documentstyle[aps]{revtex}
\draft
\title{
How to reduce the suspension thermal noise in LIGO without
improving the $Q$'s of the pendulum and violin modes.
}
\author{ V. B. Braginsky$^1$, Yu. Levin$^2$ and S. P. Vyatchanin$^1$}

\address{$^1$Physics Faculty, Moscow State University, Moscow Russia}
\address{$^2$Theoretical Astrophysics, California Institute of
Technology, Pasadena, California 91125}
\date{\today}
\begin{document}
\maketitle
\begin{abstract}

The suspension noise in
interferometric gravitational wave detectors
is caused by losses at the
top and the bottom attachments of each suspension fiber.
We use the Fluctuation-Dissipation theorem to argue
that by careful positioning of
the laser beam spot on the mirror face it is
possible to reduce the contribution of the bottom attachment point
to the suspension noise by several orders of magnitude.
For example, for the initial and enhanced
LIGO design parameters
(i.e. mirror masses and sizes, and suspension fibers' lengths and diameters)
we  predict a reduction of $\sim 100$ in the ``bottom" spectral
density throughout
  the band $35-100\hbox{Hz}$ of serious thermal noise.

We then propose a readout scheme which suppresses the
suspension noise contribution of the top attachment point.
The idea is to monitor an averaged horizontal displacement
of the fiber of length $ l$; this allows one to
record the contribution of the top attachment point to the suspension
noise, and  later subtract it from the interferometer
readout.  This method will allow a suppression factor in spectral
density of $7.4\ (l/d^2)\ \sqrt{Mg/\pi E}$, where $d$
is the fiber's diameter, $E$ is it's Young modulus and $M$ is the
mass of the mirror. For the test mass
parameters of the initial and enhanced
LIGO designs this reduction factor is
$132\times(l/30\hbox{cm})(0.6\hbox{mm}/d)^2$.

We offer what we think
might become a practical implementation of such a readout scheme.
We propose to position a thin optical waveguide close to a fused silica
fiber used as the suspension fiber. The waveguide itself is at the surface
of a solid fused silica slab which is attached rigidly to
the last mass of the seismic isolation stack (see Fig. 5).
The  thermal motion of the suspension fiber is recorded through
the phaseshift of an optical wave passed through the waveguide.
A laser power of $1\hbox{mW}$ should be sufficient to achieve
the desired sensitivity.
\end{abstract}

\pacs{ }
\section{Introduction}

Random thermal motion will be the dominant noise source
in the frequency band of $35-100$ Hz for the first
interferometers \cite{ligovirgo}
and in the frequency band of $25-126$ Hz for
the enhanced interferometers \footnote{To be specific, we refer
to the step $4$ of LIGO enhancement --- see \cite{enhancement}.
In these the suspension thermal noise was calculated assuming
the structural damping mechanism. However, the nature of
dissipation in fused silica
(e.g. viscous vs structural) is not yet fully
established for the above frequency bands.}
in  the Laser Interferometer Gravitational Wave Observatory
(LIGO)\ \footnote{The analysis of this paper is fully applicable to all other
Interferometric Gravitational Wave detectors (e.g. VIRGO, GEO-600,
TAMA etc.). For the sake of brevity in this paper we will
 refer only to LIGO.}.

The thermal noise in this frequency
band is caused by the losses in the suspension fibers, in particular
at the top and the bottom of each fiber's attachment point.
So far the only known way to reduce the thermal noise
has been
to improve the quality of the suspension fibers and their attachments.
Here  we suggest a  different approach:

In Section II we will present a general
analysis of the suspension noise based on a direct application
of the Fluctuation-Dissipation theorem. We will explicitly
separate the contributions to the thermal noise of the top
and the bottom attachment points of the suspension fibers.
It has been a common opinion that the top and
bottom attachments contribute equally to the thermal noise.
We shall challenge this point of view. In fact, we will show that
if one shifts the laser beam spot down from the center of the mirror
by an appropriately chosen distance $h$,
the contribution of the bottom attachment
point to the thermal noise can be reduced by several orders of magnitude.
Fig. 3 presents plots of this reduction factor
in the frequency band  $35$--$100\hbox{Hz}$ for three
different choices of $h$. What
is plotted here is the ratio $S_{\rm bottom}(f)/S_{\rm top}(f)$,
where $S_{\rm bottom}(f)$ and $S_{\rm top}(f)$ are the  spectral
densities of thermal noise
contributed by the bottom and the top attachment points respectively.
All three values of $h$ are close to
\begin{equation}
h={I\over M (R+l)}
\end{equation}
[cf. Eq(\ref{eq:h})],
where $l$  is the length of the
suspension fiber, $I$ is the test-mass moment of
 inertia for rotation about the center of mass in the plane of Fig. 1
(see later),
$R$ is the radius of the mirror face and $M$ is the mass of the test mass.
The numerical values of these parameters for the initial
and enhanced LIGO interferometers are

\begin{equation}
M=10\hbox{kg},\quad l=30\hbox{cm},\quad
R=12.5\hbox{cm},\label{eq:parameters}
\end{equation}
$$
I=4.73\times 10^5\hbox{g cm}^2,\quad
h=1.11\hbox{cm}.
$$
Out of the three graphs presented in Fig. 3 the one with $h=1.0\hbox{cm}$
seems to be the optimal one.
From the graphs we see that reduction factors of $\simeq 10^{-2}$ in
the ``bottom" component of the thermal noise is possible over the entire
band of serious thermal noise: $35$ to $100$ Hz.

In Sec. IIIA we   concentrate on the top attachment point. Lossy defects
at the top  create
noise not only in the test mass motion, but also noise in the motion of
the fiber.
 The latter is significantly larger than the former --- by a factor of order  $f^2
/f_{\rm pendulum}^2$
 at frequencies above
the pendulum  frequency and below the violin resonances (which  are
the frequencies
of interest for LIGO thermal noise).  We show that if one monitors
the average horizontal
displacement of
 the suspension fiber of  length $l$, one can essentially
record the fluctuating ``driving force" originating at the suspension top, and
then subtract it from
the interferometer's readout,  thereby reducing
 thermal noise originating at the
suspension top.
The reduction factor in the  spectral density
of thermal noise is given by $P=0.93\cdot l/\lambda$ [cf. Eq(
\ref{redfactor})]. Here
\begin{equation}
\lambda=
(d^2/8)\sqrt{\pi E/Mg}
\label{eq:lambda}
\end{equation}
 is the length of the
segment of fiber near it's top where
the bending is greatest, $d$ is the fiber's diameter,
$E$ is the fiber's
Young modulus and $g$ is the acceleration of gravity.
 For a fused silica fiber of  diameter
$d=0.6\hbox{mm}$ one gets a thermal noise reduction factor of
$P\simeq 132$.

In Sec. IIIB we  offer a particular way of implementing such a procedure.
The basic idea is shown in Fig. 5. A fused silica slab is rigidly attached
to the ``ceiling" (i.e. to the last mass of the seismic isolation stack),
and a
waveguide $ab$ is carved into the slab's
surface. A monochromatic optical wave
is set up in the waveguide, and a fused silica fiber used as the suspension
fiber
is positioned close to the waveguide,  within the optical wave's
evanescent field.  When the fiber is displaced relative to the waveguide,
it will change the optical wave's propagation speed, thus inducing
an overall phaseshift of the wave. The detailed calculations in Sec. IIIB
show that $\sim 1\hbox{mW}$ of optical power in the wave is sufficient to
reach
 the required
sensitivity.

\section{How to reduce thermal noise originating at the bottom
attachment point}

\subsection{The model and formalism}
The particular suspension that we consider is sketched in Fig. 1.
We consider  a compact rigid test mass of mass $M$  suspended
by a single fiber of length $l$ and mass $m$; the fiber's bottom end
is attached, for concreteness, to the top of the test mass (the main
conclusions of this paper are also valid when the test mass is suspended
by a fiber loop, as is planned for LIGO).

References \cite{gonzalez}, \cite{auld}, \cite{gusev},
\cite{levin} give detailed explanations of how
to use the Fluctuation-Dissipation
 theorem directly (without normal-mode decomposition)
to calculate the  spectral density of thermal noise \footnote{The
original formulation of the Fluctuation-Dissipation
 theorem is given in \cite{callen}}. In what
follows we
use the approach elaborated in  \cite{levin}.

To calculate the spectral density $S_{\rm x}(f)$   of suspension's
  thermal noise  at frequency
$f$ we imagine applying an oscillating force $F$
perpendicular to the test mass's
mirror surface at the center of the readout laser
beam spot \footnote{This prescription is only valid when the test masses are
perfectly rigid, which is a good approximation when dealing
with  suspension thermal noise. The case when the test masses are no
longer
considered to be rigid (e.g. for an internal thermal noise calculations)
is treated in detail in \cite{levin}. In that case the force $F(t)$
must be spread out over the laser beam spot instead of applied to
it's center point.}:
\begin{equation}
F(t)=F_0 \cos(2\pi f t).
\label{eq:force}
\end{equation}
Then $S_{\rm x}$ is given by [cf. Eq (3) of \cite{levin}]
\begin{equation}
S_{\rm x}(f)={2 k_{\rm B} T\over \pi^2 f^2}{W_{\rm diss}\over F_0^2},
\label{eq:Sx}
\end{equation}
where $W_{\rm diss}$ is the average power dissipated in the system
(suspension, in our case) when the force $F(t)$ is applied, $k_{\rm B}$
is  Boltzmann's constant and $T$ is the temperature.

For concreteness, assume that the dissipation in the fiber occurs
through  structural damping (our conclusions will hold equally well
for viscous or thermoelastic damping). In this case, the average
power dissipated during the oscillatory motion of frequency $f$ is given
by \cite{saulson}
\begin{equation}
W_{\rm diss}=2\pi f U_{\rm max}\phi(f),
\label{eq:wdiss}
\end{equation}
 where $U_{\rm max}$ is the energy of the fiber's elastic deformation
at a moment when it is maximally
bent under the action of the oscillatory force
in Eq. (\ref{eq:force}), and $\phi(f)$ is the ``loss angle" of the
material. The energy of the fiber's elastic deformation is
given by
\begin{equation}
U={JE\over 2}\int_0^l dz\left[y^{\prime\prime}\right]^2,
\label{eq:elasten}
\end{equation}
where $E$ is the Young modulus of the fiber material,
$J$ is  the geometric moment of inertia of the fiber
(for a fiber with circular cross section of diameter $d$ one has
$J=\pi d^4/64$), z is distance along the fiber
 with $z=0$ at the top and $z=l$ at the bottom,
and $y(z)$ is the fiber's horizontal displacement from
a vertical line.

This method of calculating thermal noise
 is useful for a qualitative
analysis of the system, as well as quantitative analysis.
 In particular, it allows one to see which part of the
suspension
fiber contributes the most to the thermal noise.
Assume, for a start, that the
laser
beam is positioned exactly in the middle of the mirror.
Then to work out the thermal noise
one has to imagine applying the oscillating force $F$ in Eq. (3) to the mirror
center;
the motion of the fiber and the mirror under the action of the
force are shown in
Fig. 2a.
Here we assume that the detection frequency $f$ (and hence the
frequency of the
applied force) satisfies
$f_p, f_{ r}<<f<<f_{v}$, where $f_{p}$, $ f_{r}$,
$f_{v}$ are the frequencies of the pendulum, rocking and  first
violin mode
respectively (this condition implies that horizontal
and rotational motion of the test mass is not affected by
the presence of the fiber, and that the fiber itself remains
straight).

From Fig. 2a it is clear that the fiber bends equally at the top and the bottom
 (we always assume
that at the attachment point the fiber  has to be normal to the surface to
which
 it is attached).
The total energy of elastic deformation is
\begin{equation}
U_0={1\over 2}Mg\lambda\alpha^2=
\frac{Mg\lambda}{2}\left({F\over M\omega^2l}\right)^2,
\label{eq:elasten1}
\end{equation}
where $\lambda=\sqrt{JE/Mg}$ is
 the characteristic length over which the fiber is bent near the
attachment points, $\omega=2\pi f$ is
the angular frequency of detection,
and $\alpha$ is the angle between the straight part of the fiber
and the vertical.

The bending of the fiber at the bottom can be avoided if one applies the
force $
F$ in Eq. (\ref{eq:force})
not at the middle of the mirror, but at some distance $h$ below the center.
In particular, we
should choose $h$ so that the
mirror itself rotates by the same angle as the fiber under the action of the
applied force; the resulting
motion is shown on  Fig. 2b. Physically this
 means that if we position our laser
beam at a distance $h$ below
the mirror center, then the bottom attachment point will not contribute  to
the
 thermal noise when $h$
is carefully chosen. This means that the overall suspension noise will be
reduced by a factor of order $2$
(in fact, more precisely, by a factor of $2(1+R/l)$,
where $R$ is the radius of the mirror and $l$ is the length
of the string, --- see later in this section).

In the rest of this
 section and Appendix A we find the general expression for
the suspension thermal noise, and we then work out
the optimal $h$ for the frequency band
of interest for LIGO. We will assume that when a periodic
oscillation of frequency $f$ is induced in the  system,  the average
power dissipated as heat in the suspension is given
by
\begin{equation}
W_{\rm diss}=f\left[\zeta_{\rm top}(f)\bar{\alpha}_{\rm T}^2+
               \zeta_{\rm bottom}(f)\bar{\alpha}_{\rm B}^2\right].
\label{eq:wdiss1}
\end{equation}
Here $\bar{\alpha}_{\rm T}$ and $\bar{\alpha}_{\rm B}$ are the amplitudes
of  oscillations of the
angles $\alpha_{\rm T}$ and $\alpha_{\rm B}$ respectively
(see Fig. 1), and $ \zeta_{\rm top}$ and $\zeta_{\rm bottom}$ are
frequency-dependent quantities characterizing dissipation at the top and
the
bottom respectively.
For the case of structural damping
\begin{equation}
\zeta_{\rm top}=\zeta_{\rm bottom}=\pi
f\phi(f) Mg\lambda,
\label{eq:lam}
\end{equation}
where $\lambda$ is given by Eq. (\ref{eq:lambda}) of
the introduction.

To compute  $W_{\rm diss}$ we need to evaluate
$\bar{\alpha}_{\rm T}$ and $\bar{\alpha}_{\rm B}$ by analyzing
the dynamics of the oscillations. This is done  in
Appendix A, see Eqs. (\ref{eq:alphaT}) and (\ref{eq:alphaB}).
 Putting these equations
 into Eq. (\ref{eq:wdiss1}) and then into Eq. (\ref{eq:Sx}), we obtain
[cf. Eq. (\ref{eq:Sx1})]
\begin{eqnarray}
S_{\rm x}(f)&=&{4 k_{\rm B} T\over\pi \omega}
 \left\{ {I/M-R(g/\omega^2+h)\over \left[Ig-MgR\left(
g/\omega^2-R\right)\right]\cos(k l)-
(I\omega^2-MgR)\sin(kl)/k}\right\}^2\nonumber\\
& &\times\left\{\zeta_{\rm top}+\zeta_{\rm bottom}\cos ^2(kl)
\left[{I/M-h\left[R+\tan \left(k l\right)/k-g/\omega^2\right]\over
I/M-R(g/\omega^2+h)}\right]^2\right\}.
\label{eq:Sx11}
\end{eqnarray}
Here $k=\omega/c=2\pi f/c$, $c=\sqrt{glM/m}$ is the speed of propagation
of
a transverse wave in the fiber.
From the above equation we can infer
the ratio of the bottom and the top
 contributions to the thermal
 noise:
\begin{equation}
{S_{\rm bottom}(f)\over S_{\rm top}(f)}={\zeta_{\rm bottom}(f)\over \zeta_{
\rm top}(f)}
\cos^2(kl)\left[{I/M-h\left[R+\tan \left(k l\right)/k-g/\omega^2\right]\over
I/M-R(g/\omega^2+h)}\right]^2.
\label{eq:ratio6}
\end{equation}
This is the most important equation in this section of the paper; it
will be discussed in the next subsection.

\subsection{The case of low-frequency suspension noise}
When the detection frequency $f$ is far below the frequency of the
fundamental violin mode, $f_{\rm v}$,  then $kl\ll 1$ in Eq. (\ref{eq:ratio6})
and
\begin{equation}
{\tan(kl)\over k}\simeq l\left[1+{1\over 3}\left(kl\right)^2\right].
\label{eq:tan}
\end{equation}
Let us assume that the top and the bottom are equally lossy, i.e.
$\zeta_{\rm top}=\zeta_{\rm bottom}$, as the would be
for structural damping, Eq (\ref{eq:lam}) above.
We choose $h$ to be

\begin{equation}
h={I\over M(R+l)}.
\label{eq:h}
\end{equation}

Putting Eqs (\ref{eq:h}) and (\ref{eq:tan}) into Eq. (\ref{eq:ratio6}), we get

\begin{equation}
{S_{\rm bottom}(f)\over S_{\rm top}(f)}\simeq {\pi^4\over 9}{1\over
\left[1-(R/h)(\omega_p^2/\omega^2)\right]^2}\left({f\over f_{\rm v}}\right)^4,
\label{eq:SbSt}
\end{equation}
where $\omega_p=\sqrt{g/l}$.

For the initial and
enhanced LIGO design $f_{\rm v}\simeq 400\hbox{Hz}$,
$M\simeq 10\hbox{kg}$, $I\simeq 4.73\times
10^{-2}\hbox{kg}\times\hbox{m}^2$,
$R\simeq 12.5\hbox{cm}$, and the interesting frequency range where
suspension noise is expected to dominate is $35-100\hbox{Hz}$
(actually, this depends on the stage of enhancement. The frequency
band specified above
is where the suspension thermal noise is expected
to dominate in the initial LIGO; in the enhanced version
this frequency interval will be larger). In this
case Eq. (\ref{eq:SbSt}) gives
$ S_{\rm bottom}(f)/ S_{\rm top}(f)\simeq 0.002-0.2$.

In Fig.3 we give plots for $S_{\rm bottom}/S_{\rm top}$ as
a function of the detection frequency $f$ for three different choices of $h$.
We have used Eq. (\ref{eq:SbSt}) to make all the plots and we set $I$,
$M$, $R$ and
$l$ to the numerical values appropriate for the initial and
enhanced LIGO design
and given at the beginning of this section.

The first curve
 is plotted for $h$ given by Eq. (\ref{eq:h}), in our case
$h=1.11\hbox{cm}   $. The second and third curves are  for
$h=1.0\hbox{cm
}$
 and $h=0.9\hbox{cm}$ ;
 these values of $h$ are chosen so that $S_{\rm bottom}/S_{\rm top}=0$
for $f=80$Hz and $f=105$Hz respectively.
Out of the three cases the choice $h=1\hbox{cm}$ gives the best overall
performance across the considered frequency band, with the typical
reduction factor of
\begin{equation}
{S_{\rm bottom}\over S_{\rm top}}\sim 10^{-2}.
\end{equation}
>From Eq. (\ref{eq:Sx11}) we see that choosing $h$ close to the value
in Eq. (\ref{eq:h}) reduces the total suspension thermal noise
by a factor close to $2(1+R/l)\sim 3$ relative to the case when
$h=0$.

\subsection{High-frequency suspension thermal noise}
A somewhat less interesting observation is that for $h=0$ and
$f_{\rm n}=f_{\rm v}(n+1/2)$, where $n$ is an integer,
\begin{equation}
{S_{\rm bottom}(f_{\rm n})\over S_{\rm top}(f_{\rm n})}=0.
\label{eq:fn}
\end{equation}
Unfortunately, at $f=f_{\rm n}$ the interferometer's noise is dominated
by  shot noise. However, if one uses an advanced optical topology --- for
example,  resonant sideband extraction --- then it is possible to reduce
the shot noise in a narrow band around any chosen frequency. Then the
thermal
noise may dominate in this narrow band, and our observation (\ref{eq:fn})
may be useful in case one tries to reduce the thermal noise by  cooling
of the fiber top.

\section{How to control noise from the top}
\subsection{The concept}

In this section we propose a recipe for how to decrease the influence
of the
 thermally fluctuating stress at the top part of the suspension fiber.
The basic idea is the following:

Intuitively, the fluctuations at the top cause bending of the fiber
at the top, which will be a random process in time.
This random bending
 will randomly move the rest of the fiber and ultimately drive
the random motion of the test mass.
We propose to measure directly
the thermally driven fluctuations in the horizontal displacement of the fiber,
and from them
 infer  the fluctuating force which drives
the random motion of the mirror. We can then subtract the motion due to
this fluctuating force from the interferometer output\footnote{The idea
of thermal noise compensation is not new (e.g. \cite{weiss},
\cite{kulagin}).
  However, our
 detailed treatment and concrete
experimental proposal is different from anything prior
to this paper.}.

Formally this amounts to introducing a new readout variable $q$
as follows:
\begin{equation}
q=X_{\rm mirror}+X_{\rm fiber}.
\end{equation}

Here $X_{\rm mirror}$ is the horizontal displacement of the laser spot's
center (i.e.the signal ultimately read by the interferometer's
photodiode), and
\begin{equation}
X_{\rm fiber}=\int_0^l dz\Phi(z)y(z)
\end{equation}
is the fiber's horizontal displacement weighted by some function
$\Phi(z)$ to be discussed below.
We will postpone the discussion of how to measure $q$ experimentally
until the next section; here we concentrate on finding  the optimal
$\Phi(z)$ and seeing what is the maximal possible reduction in the
thermal noise.

To find the spectral density of fluctuations in $q$ we need to imagine
acting on
the system with sinusoidal force $F_q\propto \cos(2\pi ft)$
that appears in the interaction
hamiltonian in the following way
\begin{equation}
H_{\rm int}=-qF_q=-X_{\rm mirror}F_q-\int_0^l dz F_q\Phi(z) y(z);
\label{eq:Fq}
\end{equation}
cf. the discussion of the Fluctuation-Dissipation theorem in
Ref. \cite{levin}.
From the Eq. (\ref{eq:Fq})
 we observe that applying the generalized force
$F_q$ to the system is equivalent to applying two  forces simultaneously:
one is a force of
magnitude $F_q$
 applied to the mirror surface at the
center of the beam spot,
and the other  is a force  distributed along the fiber
in the following manner:
\begin{equation}
{dF_{\rm fiber}\over dz}=F_q\Phi (z).
\end{equation}
The resulting motion of the system is shown in Fig. 4. The intuitive
idea is to choose the weighting function
$\Phi(z)$ so that when the beam spot's
height $h$ has also been appropriately
chosen, $F_q$ induces no
bending of the fiber at the top or at the bottom.

In the case of  structural damping the
dissipated power  is
proportional to the elastic energy $U$ of the fiber. Thus formally
one has to
choose $\Phi(z)$ and $h$ so that $U$ is minimized. It is convenient to
reformulate the problem: to find the shape of the fiber $y(z)$ and
beam-spot height $h$ for
which the functional in Eq. (\ref{eq:elasten}) has a minimum, and after this
calculate the distribution
$\Phi(z)$ of the driving force on the fiber
that will produce the desired shape $y(z)$.
In Appendix B we carry out this straightforward but somewhat tedious
task. We obtain [cf. Eq. (\ref{eq:B3})]
\begin{eqnarray}
y_{\rm optimal}(z)&=&{F_q\over M\omega^2 }\left({z\over
                     l}\right)^2\left({3(r+1)-z/l\over
                     2(3r^2+3r+1)}\right)\nonumber\\
                  & &\simeq{F_q\over M\omega^2 }\left({z\over l}\right)^2
                       \left(0.76-0.18{z\over l}\right).\label{eq: yopt}
\end{eqnarray}
Here $r=R/l$, $R$ is the radius of the mirror, $l$
is the length of the fiber, $\omega=2\pi f$ is the angular frequency
of detection.
We substitute
                 here and below $r=0.42$ corresponding to the initial
and enhanced LIGO test masses. The profile of the distributed force acting
on the fiber
and hence of $\Phi(z)$ is mainly determined by $y^{\prime\prime}(z)$ (see
Appendix B):
\begin{equation}
\Phi(z)\simeq \Phi_0= -{Mg\over F_q}y^{\prime\prime}(z),
\end{equation}
which gives [cf. Eq. (\ref{eq:Blast})]
\begin{eqnarray} \label{phi0}
\Phi_0(z)&\simeq& -{\omega_p^2\over \omega^2 l}\left(1+r-{z\over l}\right)
                             {3\over 3r^2+3r+1}\nonumber\\
                     &=&-{\omega_p^2\over \omega^2 l}\left(1.53-1.08{z\over
                                       l}\right),\label{eq:phiopt}
\end{eqnarray}
where $\omega_p=\sqrt{g/l}$.
When the force distribution has this optimal form,
the elastic energy has the minimum value
\begin{eqnarray}
U_{\rm min}&\simeq&{3\over 3r^2+3r+1}{\lambda\over l}{Mg\lambda\over 2}
                   \left(F_q\over M\omega^2 l\right)^2\nonumber\\
           &=&{1.08\lambda\over l}\times U_0,\label{eq:umin}
\end{eqnarray}
where $U_0$ is the elastic energy in Eq. (\ref{eq:elasten1}). Therefore, for
a fused
silica fiber with $E\simeq 6.9\times 10^{10}\hbox{Pa}$ and
$d=0.6\hbox{mm}$, we get $\lambda\simeq 2.1\hbox{mm}$ and the
maximal
reduction factor for the spectral density
of suspension thermal noise is
\begin{equation}\label{redfactor}
P=\frac{l}{1.08\lambda}\simeq  132.
\end{equation}

\subsection{Experimental realization--a proposal}
\subsubsection{Preliminary remarks}
Before describing a particular experimental realization
of the above scheme,
a few general remarks are in order.

First, one might worry that our averaging function $\Phi(z)$ is
frequency  dependent --- in general, that could make the experimental
implementation very difficult. In particular [see Appendix B,
Eq.(\ref{phi01})], $\Phi$ consists of two components:
$\Phi=\Phi_0+\Phi_1$, where $\Phi_0$
and $\Phi_1$ as given by Eq. (\ref{phi01})
have very different frequency dependence.
However at the frequencies
of interest $\Phi_0 \gg \Phi_1$, and  then the approximate formula
  (\ref{eq:phiopt}) for the averaging function $\Phi(z)=\Phi_0(z)$ is a
product of  two terms:  one  which depends only on the frequency
$f$ (i.e.  $\Phi(z)\propto 1/f^2$), and the other which
 depends only on the
coordinate $z$. This feature makes the scheme feasible for a broad
range of frequencies.  It is sufficient that our device measures the
displacement of the fiber with the frequency-independent averaging
function $\tilde{\Phi}(z)\propto f^2\times \Phi(z)$, and
that the frequency dependence is then
put back in during data analysis
when constructing the readout variable
$q$:
\begin{equation} q=X_{\rm mirror}+\eta(f)\int_0^l dz
\tilde{\Phi}(z)y(z), \end{equation}
 where $\eta(f)\propto f^{-2}$ is chosen so that
$\eta\tilde{\Phi}=\Phi$.

As mentioned above, Eq. (\ref{eq:phiopt}) is an approximation valid
when the fiber has no inertia, i.e. when
$f\ll f_v=\hbox{(lowest violin-mode frequency)}$. When the inertia
of the fiber becomes important ($\Phi_1\sim \Phi_0$),
it is no longer possible to factor out
a frequency-dependent part of $\Phi$. As a result, when $f$ gets
closer to $f_v$, the effectiveness of the thermal noise suppression
(i.e. the value of $P$) is reduced. A detailed
 analysis shows that if
we choose $\tilde{\Phi}(z)$ so that the thermal noise compensation is
optimal ($P=P_{\rm max}$) at low frequencies $f\ll f_v$, then at $f=0.2
f_v$ we have $P\sim 0.9P_{\rm max}$, at $f=0.32f_v$ we have $P\sim 0.5
P_{\rm max}$, and beyond this $P$ is reduced sharply as we
approach the first violin mode. For the fused
silica fiber discussed above $f_v\sim
400\hbox{Hz}$, so the compensation is
effective throughout the band $35-100$Hz where suspension thermal
noise dominates. It is worth  emphasizing that this deterioration
in the reduction factor only happens when
we use the averaging function $\Phi_0$ instead of
$\Phi_0+\Phi_1$ close to the violin frequency.  Thus, this limitation
is one of technology and not of principle. Perhaps, it is possible
to conceive of a   scheme where the correct averaging function
is implemented at all frequencies. However, we have not
been able to do so.

Secondly, any sensor used for monitoring the fiber coordinate $X_{\rm
fiber}$ will have an intrinsic noise which will deteriorate the quality
of the thermal-noise compensation.
In particular, the  overall reduction factor $P_{\rm eff}$
is given by
\begin{equation}
{1\over P_{\rm eff}}={1\over P}+{S_{ \rm fiber\
meas}(f)\over S_{\rm fiber\ therm}(f)},
\end{equation}
where $S_{ \rm fiber\ meas}(f)$ is the spectral density
of  intrinsic noise of the
device which measures  the
average displacement
of the fiber  and $S_{\rm fiber\ therm}(f)$ is the spectral density of
thermal fluctuations of the same displacement.

For the case of structural damping it is easy to estimate
\begin{equation}
\sqrt{S_{X\ \rm fiber\ therm}(f)f}\sim\sqrt{\frac{\lambda\phi kT}{Mg}
}\sim
10^{-14}\hbox{cm},
 \end{equation}
where we assume that $\phi\sim 10^{-7}$ for fused silica.
If our goal is to achieve $P\sim 100$ then the condition $P_{\rm
 eff}\simeq P$ implies
\begin{equation}
\sqrt{S_{ \rm fiber\ meas}f}<<\sqrt{S_{\rm fiber\ therm}f\over P}\sim
 10^{-15}\hbox{cm}.
\end{equation}
We shall take the above number as a sensitivity goal that
our measuring
device should achieve.

\subsubsection{Proposed measuring device}
Now we are ready to describe a possible practical implementation
of our thermal-noise compensation scheme.
Figure 5 illustrates the basic idea.
We propose to use a fused silica optical fiber with the
refractive index $n_1$  for the test mass's suspension.
 Next to this fiber
 we attach to the top seismic isolation plate
(i.e. the ``ceiling") a rigid block of the fused silica $A$ with
the same index of refraction $n_1$. On the surface
of this rigid block we put  a thin optical
waveguide with refractive index $n_2$ such that
$n_2>n_1$, so that the waveguide is
at a distance $\sim \lambda_{\rm optical}/2\pi$ from the suspension fiber.
It is assumed that the side of the waveguide close
to the suspension fiber does not have any coating,
i.e. it is ``naked".
In this configuration the optical wave may propagate through the
waveguide without  substantial scattering even though the suspension fiber
is within the wave's evanescent  zone. This device will produce a
relatively  large response to the displacement $X_{\rm fiber}$ in the
form of a phaseshift of $\Delta \phi$ of the optical wave:
\begin{equation}
\Delta \phi=K{2\pi X_{\rm fiber}\over \lambda_{\rm optical}}{2\pi l\over
\lambda_{\rm optical}},
\label{eq:pshift}
\end{equation}
where the dimensionless factor $K$ depends on the values of $n_1$ and
$n_2$ and for  typical optical waveguides is $K\sim 10^{-3}$.
 Equation (\ref{eq:pshift}) implies that
 in order to register $X_{\rm fiber}\sim
10^{-15}\hbox{cm}$ we need a sensitivity $\Delta\phi\sim
10^{-7}$. Thus for averaging time of
$\tau_{\rm grav}=0.01$sec we need to use the power
of coherent light of $W\sim
1\hbox{mW}$. This power can be
decreased if one uses a resonant standing wave in the waveguide.

Apart from the shot noise of the laser light, let us briefly discuss
two other kinds of noise in this sensor. A more complete discussion
will be presented elsewhere.

The first kind is  seismic noise. A simple calculation shows that
the seismic contribution to the noise in the readout
variable $q$ is about twice as large in spectral density as the seismic
contribution to the
noise in $X_{\rm mirror}$. Thus the seismic noise will not be an issue
at frequencies above the ``seismic wall'' of the LIGO sensitivity curve.

The second kind of noise we want to mention is the mechanical
thermal fluctuations of the waveguide itself.  Our estimates show
that if these fluctuations are caused by  structural damping (and not
by some surface or contact defects), then the ratio of the mechanical
thermal fluctuations of the waveguide to those of the fiber is
\begin{equation}
{S_{\rm waveguide}\over S_{\rm fiber}}\sim {Mg\over El\lambda}\sim
10^{-5}.
\end{equation}
Thus, if the system is sufficiently clean then the mechanical thermal
fluctuations of the waveguide will probably
not significantly reduce the sensitivity
of our sensor.

It is worth noting that in order
to achieve the optimal compensation of thermal noise,
the distance $d(z)$ between the suspension
fiber and the waveguide  has to vary in accord with the
optimal profile of the averaging function:
\begin{equation}
d=A-B\log\left[\Phi(z)\right],
\end{equation}
where $A$ and $B$ are constants to be discussed elsewhere.
In this case the phase of the waveguide's output records the
optimally averaged coordinate $X_{\rm fiber}$ of the fiber.

The profile $d(z)$ may be difficult for experimental realization. However we
find that in the simplest case when $\Phi(z)$ is a constant over
the length $l$ of
averaging, the factor $P$ is reduced very little: from $P=132$ to $P\sim
120$.

 \section{Conclusion}
 In this
paper we have done two things.

Firstly, we have shown that by an
appropriate positioning of the laser's beam spot on the  surface
of each test-mass mirror, one
can reduce the contribution of  the suspension fiber's bottom to the
suspension thermal noise by two to three orders of magnitude in the
frequency band of $35-100\hbox{Hz}$ for the initial LIGO design.

Secondly, we have proposed a way to compensate the suspension
thermal noise originating from the top of
each fiber by monitoring independently
the fiber's random horizontal displacement. In the best case,
with the system parameters for the initial or enhanced
 LIGO design, one can get
a reduction factor of the order of $P=130$ in spectral density
over the entire $35-100$Hz band,
when both the first and second procedures are applied;
and with realistic defects in the design one should be
able to get a reduction of at least $P\simeq 100$

The device that compensates the suspension thermal noise
can ease the requirements to quality of suspension system.
In particular, if this device allows the reduction factor
of $P=100$, this would effectively increase the quality
factors of pendulum and violin modes by a factor of $P=100$.
So far the highest quality factor $Q\simeq 10^8$ of the pendulum mode
was achieved in \cite{braginsky} for a fused silica suspension
fiber, which allows one to reach
the Standard Quantum Limit for
averaging time of $10^{-3}$sec.
 Implementation of our proposal could effectively
increase this quality factor to $Q_{\rm eff}\simeq 10^{10}$,
which would reduce the thermal noise in LIGO to the level of
Standard Quantum Limit for averaging time
of $10^{-2}$sec. Then the techniques which allow one
to beat the Standard Quantum Limit (see e.g. \cite{QND}) could
be used in the enhanced LIGO interferometers.

\section*{acknowledgments}
 We thank  Sergey Cherkis, Michael Gorodetsky, Ronald Drever, Viktor
Kulagin, Nergis Mavalvala, Peter Saulson and Kip Thorne for interesting
discussions. We are grateful to Kip Thorne for carefully looking over
the manuscript and making many useful suggestions. This research has
been supported by NSF grants PHY-9503642 and PHY-9424337, and by
the
Russian Foundation for Fundamental Research grants \#96-02-16319a and
\#97-02-0421g

\section*{Appendix A}
In this appendix we solve the
 dynamical problem of finding the amplitudes $\bar\alpha_{\rm T}$
and $\bar\alpha_{\rm B}$ of oscillation of the top and
bottom bending angles in Eq. (\ref{eq:wdiss1}) when a periodic force
\begin{equation}
F=F_0\cos(\omega t)
\end{equation}
is applied to the mirror at a distance $h$ below the mirror center
[we use these amplitudes in Eq. (\ref{eq:wdiss1}) of the text].
For convenience we complexify all of the quantities:
$$
F=F_0 e^{\imath \omega t},\quad
\alpha_{\rm T}=\bar{\alpha}_{\rm T} e^{\imath  \omega t},\quad
 \alpha_{\rm B}=\bar{\alpha}_{\rm B} e^{\imath \omega  t},
$$
$$
 x=\bar{x} e^{\imath \omega t},\quad
\psi=\bar{\psi}e^{\imath \omega t},$$
where $x$ is the  displacement of the test mass's center of mass
and $\psi$ is the angle by which the mirror is rotated (see Fig. 1)
under the action
of the force $F(t)$. As usual, $\omega=2\pi f$ is the angular frequency.

>From the projection of the  Newton's Second Law on the horizontal
axis we have
\begin{equation}
F_0-(\bar{\alpha}_{\rm B}-\bar{\psi}) M g=-M\omega^2\bar{x},
\label{eq:newton}
\end{equation}
and, for the rotational degree of freedom, the equation of motion is
\begin{equation}
F_0h+MgR\bar{\alpha}_{\rm B}=I\omega^2\bar{\psi},
\label{eq:newton1}
\end{equation}
where $R$ is the radius of the test-mass cylinder and $I$ is the moment
of inertia for  rotation about the test-mass center of mass in the
plane of the Fig.\ 1.
In the two equations above  we assume that $\alpha_{\rm B}$ and
$\psi$ are small.

The fiber's horizontal
displacement $y$ from a vertical line approximately satisfies
the wave equation:
\begin{equation}
{\partial^2 y\over \partial t^2}=c^2{\partial^2 y\over\partial z^2},
\label{eq:wave1}
\end{equation}
where $z$ is distance along the wire, with
$z=0$ at the top and $z=l$ at the bottom, and $c=\sqrt{glM/m}$ is the
transverse speed of sound in the wire.
In this Appendix we use Eq. (\ref{eq:wave1}) for
flexible wire since it's solutions are simple.
If one takes the stiffness into account this changes the  solutions of
Eq.\ (\ref{eq:wave1}) by a relative order of $\lambda/l$, see e.g.
\cite{blandford}. However, when using Eq. (\ref{eq:wave1}),
we must allow non-zero bending angles at the top and bottom attachment
points,
 $\alpha_{\rm T}$ and $\alpha_{\rm B}$. The energy of elastic strain
of the wire then consists of two components: one from the bulk
of the wire given by Eq. (\ref{eq:elasten}), and the other from
the bending at the attachment points given by Eq. (\ref{eq:elasten1}).
The solution to Eq. (\ref{eq:wave1}) is

\begin{equation}
y(z,t)=A\sin\left({k z}\right) e^{\imath \omega t},
\label{eq:y}
\end{equation}
where $k=\omega/c$ is the wave vector of an off-resonance
standing wave induced in the fiber and $A$ is a constant.
The boundary condition is set at the bottom by
\begin{eqnarray}
A\sin\left({k l}\right)&=&\bar{x}+R\bar{\psi}\nonumber\\
{k A}\cos\left({k l}\right)&=&\left(\bar{\alpha}_{\rm B}-\bar{\psi}
\right).\nonumber
\end{eqnarray}
Putting these two equations into Eqs. (\ref{eq:newton}) and
(\ref{eq:newton1}),
 we find
\begin{equation}
\bar{\alpha}_{\rm B}=-F_0{I/M-h\left[R+\tan \left(k
l\right)/k-g/\omega^2\right]\over
 MgR^2+(I\omega^2-MgR)\left[g/\omega^2-\tan \left(k l\right)/k\right]}
\label{eq:alphaB}
\end{equation}
and
\begin{equation}
\bar{\alpha}_{\rm T}=-F_0{I/M-R(g/\omega^2+h)\over \left[Ig-MgR\left(
g/\omega^2-R\right)\right]\cos(k l)-(I\omega^2-MgR)\sin(kl)/k}.
\label{eq:alphaT}
\end{equation}
Putting Eqs. (\ref{eq:alphaB}), (\ref{eq:alphaT})
 and (\ref{eq:wdiss1}) into Eq.\ (\ref{eq:Sx}),
we finally get for the spectral density of the suspension thermal noise:
\begin{eqnarray}
S_{\rm x}(f)&=&{4 k_{\rm B} T\over\pi \omega}
 \left\{ {I/M-R(g/\omega^2+h)\over \left[Ig-MgR\left(
g/\omega^2-R\right)\right]\cos(k l)-
(I\omega^2-MgR)\sin(kl)/k}\right\}^2\nonumber\\
& &\left\{\zeta_{\rm top}+\zeta_{\rm bottom}\cos ^2(kl)
\left[{I/M-h\left[R+\tan \left(k l\right)/k-g/\omega^2\right]\over
I/M-R(g/\omega^2+h)}\right]^2\right\}.
\label{eq:Sx1}
\end{eqnarray}

\section*{Appendix B}

Here we calculate the optimal shape $y_{\rm optimal}(z)$
of the fiber and the vertical
position of the laser beam spot $h$ that minimize the
fiber's elastic deformation energy [Eq.\ (\ref{eq:elasten})].

It is easy to deduce from Eq. (\ref{eq:elasten})
 that energy minimizing function $y(z)$ obeys the equation
$y^{\prime\prime\prime\prime}(z)=0$.  Therefore

\begin{equation}
  \frac{y(z)}{l}=a_0+a_1\frac{z}{l}+a_2\frac{z^2}{l^2}+a_3\frac{z^3}{l^3},
\label{eq:as}
\end{equation}
where $a_i$ are  constants to be determined.

Let us discuss the boundary conditions.
Strictly speaking, the boundary conditions should be such that the fiber
is perpendicular to the surface of attachment at both the top and the
bottom.
Therefore at the top we have $y(0)=y^{\prime}(0)=0$, from which
immediately follows $a_0=a_1=0$.
However, at the bottom
it is more convenient for our calculations
embody the bending of the fiber, on the lengthscale
$\lambda$, in a bending angle $\alpha_{\rm B}$ as in Fig.\ 1,
and correspondingly add an additional term
\begin{equation}
U_{\rm add}=(1/4)Mg\lambda\alpha_{\rm B}^2
\label{eq:uadd}
\end{equation}
 to the energy functional in
Eq. (\ref{eq:elasten}), and then in Eq. (\ref{eq:as})
evaluate $y(l)$ and it's derivatives above
the $\lambda$-scale bend.
Our energy minimization procedure will make the
angle $\alpha_{\rm B}$
so small that the additional elastic energy as given
by
Eq. (\ref{eq:uadd})
 is negligible compared to $U$ in
Eq. (\ref{eq:elasten})

The coefficients $a_2$ and $a_3$ can be inferred from
force and torque balance at the test mass:
\begin {equation}
F_q - Mg y'(l)=-M\omega^2( y(l) +R(y'(l) + \alpha_{\rm B})),
\end {equation}
and
$$
F_q h-MgR\alpha_{\rm B}= -I\omega^2(y'(l)+\alpha_{\rm B}).
$$

It is useful to rewrite these equations in a dimensionless form:

\begin {eqnarray} \label{eq:dyneq}
\xi (1 + \eta (r - a)) &+& r \alpha_{\rm B}=- \xi_0,
\nonumber\\
\eta \xi &+&\alpha_{\rm B} (1-\mu r a)=-\mu s \xi_0;
\end{eqnarray}
where
$$
\xi=\frac{y(l)}{l},\ \ \eta=\frac{y'(l)l}{y(l)}, \ \ s=\frac{h}{l},$$
$$
a=\frac{\omega_{p}^2}{\omega ^2}\simeq 10^{-3} \ \div 10^{-6 }, \ \
r=0.42, \ \ \mu =\frac{Ml^2}{I}=19,
$$
where $\omega_p=\sqrt{g/l}$.
Here we have used for estimates the mirror parameters for
the initial and enhanced LIGO interferometers.
Solving the above system
of equations (\ref{eq:dyneq}) for  $\xi$ and $\alpha_{\rm B}$
  (taking $\eta$
as a parameter) we get:

$$
\alpha_{\rm B} =\xi_0 \frac {\eta - \mu s (1+\eta(r-a))} {[1+\eta(r-a)][1-\mu
ra]-r\eta}
\simeq \xi_0[\eta-\mu s (1+\eta (r-a))]
$$
$$
\xi=-\xi_0 \frac{1-\mu r(a+s)}{[1+\eta(r-a)][1-\mu ra]-r\eta}
\simeq -\xi_0 (1-\mu r(a+s))
$$
Let us choose the parameter $s$ so that $\alpha_{\rm B}=0$ for
some angular frequency $\omega_0$ in the  frequency band $35-100$Hz
where thermal noise is most serious:

\begin{equation}\label{s}
s\simeq \frac{\eta}{\mu[1+\eta(r-a_0)]} \simeq \frac{\eta}{\mu[1+\eta r]} , \ \
a_0=\frac{\omega_{p}^2}{\omega_0^2}
\end{equation}
Then we get for $\alpha_{\rm B}$ and $\xi$
\begin {equation} \label {beta}
\alpha_{\rm B} \simeq \xi_0 \frac{\eta^2}{1+\eta r}(a-a_0)
\end {equation}
$$ \xi\simeq -\xi_0 \frac{1}{1 + \eta r}.$$

We can express the coefficients $a_3$ and $a_2$ in terms of $\xi$ and
$\eta$
by combining Eqs. (\ref{eq:as}) and (\ref{eq:dyneq}),
and we can then
calculate the elastic energy  according to Eq. (\ref{eq:elasten}):

\begin {equation}
U\simeq \frac {Mg \lambda}{2} \left(\frac{F_q}{M\omega^2 l}\right)^2
\times
\frac{\lambda}{l}\times  \frac{4(\eta^2-3\eta+3)}{(1+r\eta)^2}
\end {equation}
This function has the minimal value
$$
U_{\rm min}\simeq \frac{ l}{\lambda} \times \frac{3}{1+3r+3r^2}
\times U_0=
\frac{1.08\lambda}{l} \times U_0
$$
at optimal $\eta$ given by
\begin{equation}
\eta_{\rm opt}=\frac{3(1+2r)}{2+3r}=1.69.
\label{eq:eta}
\end{equation}
Here $U_0$ is the energy of elastic strain of the fiber
when the force of magnitude $F_q$ is applied in mirror
center, as worked out in Eq. (\ref{eq:elasten1}).
Now we can figure out the optimal shape of the fiber's horizontal
displacement:
\begin{eqnarray}
y_{\rm optimal}(z)&=&{F_q\over M\omega^2 }\left({z\over
                     l}\right)^2\left({3(r+1)-z/l\over
                     2(3r^2+3r+1)}\right)\nonumber\\
                  & &\simeq{F_q\over M\omega^2 }\left({z\over l}\right)^2
                       \left(0.76-0.18{z\over l}\right).\label{eq:B3}
\end{eqnarray}
From  Eq. (\ref{s}) we get $h=l\times s\simeq 1.55\hbox{cm}$.

Using (\ref{beta}) one can
show that $\alpha_{\rm B} \le 1.7\cdot 10^{-3}
\cdot \xi_0 $ over the frequency
band $35-100$Hz. From this and Eq.\ (\ref{eq:uadd}),
one can compute the energy due to the bending at the
fiber bottom:
$U_{\rm add}\simeq 1.4\cdot
10^{-6}\times E_0$. We see that $ U_{\rm add} \ll U_{\rm  min}$
and hence over the frequency band of interest the small
bending at the bottom  does not contribute significantly to the
total energy of elastic deformation.

The profile of the distributed
force and correspondingly the function $\Phi$ are
given by

\begin {equation} \label{phi}
F_q \Phi (z)= -\rho \omega^2 y(z)  - Mg y''(z) +IE
y''''(z).
\end {equation}
Here $\rho$ is the fiber density per unit length.
Since $y^{\prime\prime\prime\prime}(z)=0$,
the function $\Phi$ consists of two terms
$\Phi(z)=\Phi_0(z)+ \Phi_1(z)$, where
\begin{equation} \label{phi01}
\Phi_0(z) = -\frac{Mg}{F_q} y''(z), \ \
\Phi_1(z) = -\frac{\rho\omega^2}{F_q} y(z).
\label{eq:Blast1}
\end{equation}
\begin{equation}
\Phi_0(z)= \frac{\omega_p^2}{l\omega^2}\cdot
\left(1+r-\frac{z}{l}\right)\cdot \frac {3}{3r^2+3r+1}
=\frac{\omega_p^2}{l\omega^2} \cdot \left(1.53-1.08 \frac {z}{l}\right).
\label{eq:Blast}
\end{equation}
We see that  $\Phi_0$ is much greater than $\Phi_1$ in our frequency
range
($10-100\hbox{Hz}$ for the initial LIGO).

\newpage

\begin{figure}
\input{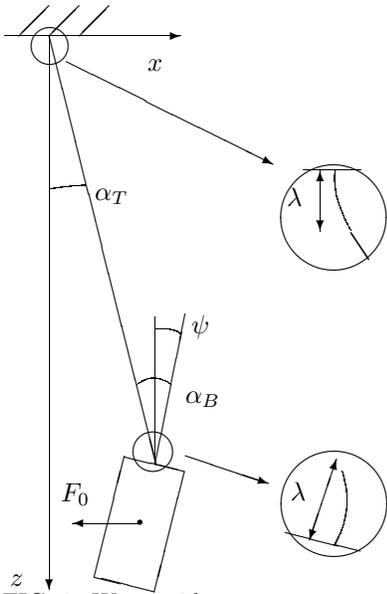}
\caption{We consider a test mass suspended on a single fiber. The fiber's
bottom
is attached to the top of the test mass, and the fiber's top is attached to the
last stage of the seismic isolation stack. It is assumed that at attachment
points
the fiber is perpendicular to the surface to which it is attached.
 }
\end{figure}

\begin{figure}
\input{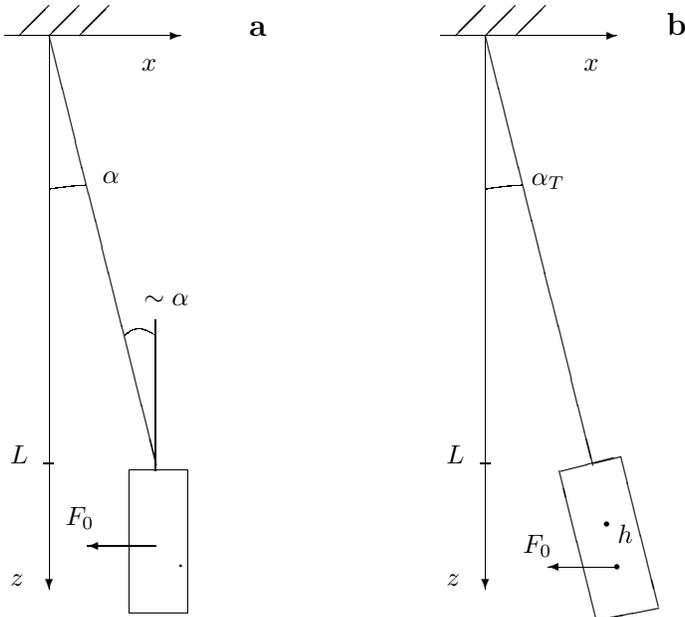}
\caption{Motion of the test mass and the suspension fiber
under the action of an oscillating force applied at the center
of the laser beam spot in two different cases: {\it a)}  the beam
spot is positioned at the mirror center, the fiber bends equally at the
top and the bottom, and {\it b)}  the position
of the beam spot is shifted down from the center of the mirror, so that
there is no bending of the fiber at the bottom.}
\end{figure}
\newpage

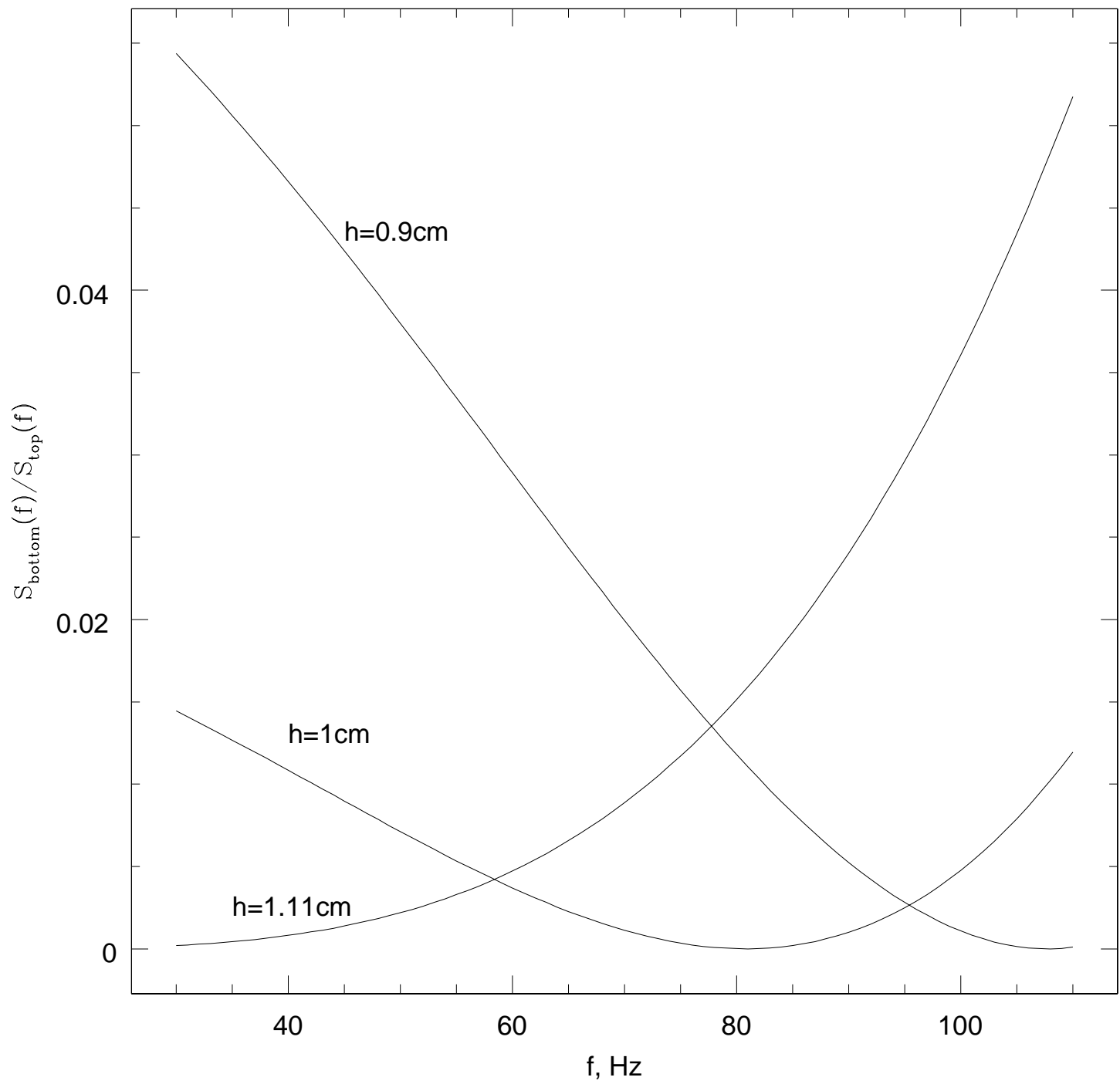
\begin{figure}
\caption{Fig. 3. 
A plot of $S_{\rm bottom}(f)/S_{\rm top}(f)$ as a function
of frequency $f$ for three different positions of the laser beam spot.}
\end{figure}

\begin{figure}
\input{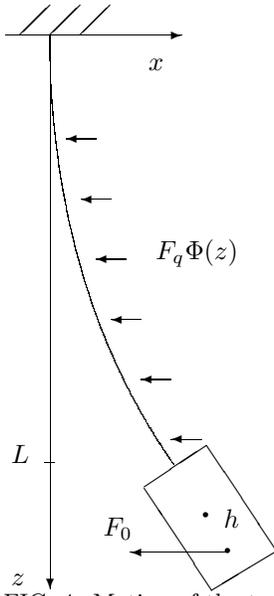}
\caption{Motion of the test mass
and suspension fiber under the action of the generalized
force $F_q$ defined in Eq. (\ref{eq:Fq}) of the text.
The force $F_q$ should be chosen so that there is no bending
of the fiber at it's top and bottom attachment points.}
\end{figure}

\begin{figure}
\input{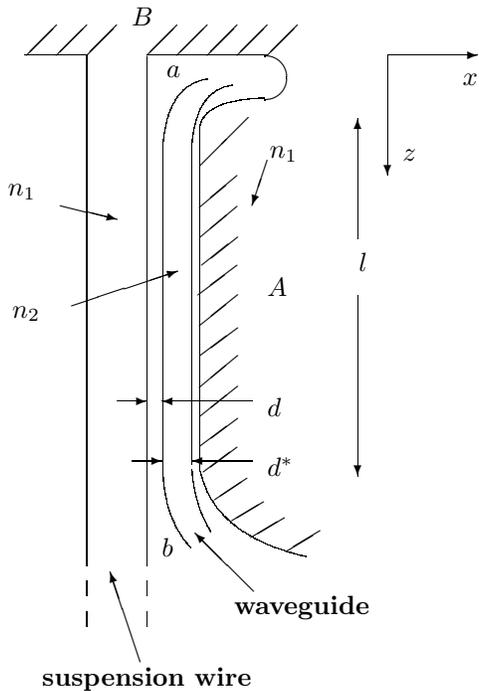}
\caption{A proposed scheme for compensation
of a suspension thermal noise. The optical waveguide $ab$ is positioned
close to the suspension fiber made of fused silica. A horizontal
displacement
of the suspension fiber is recorded through a phase shift of
an optical wave propagating through the waveguide.}
\end{figure}

\end{document}